\documentclass[english]{article}
\usepackage{lmodern}

\usepackage[T1]{fontenc}
\usepackage[latin9]{inputenc}
\usepackage{geometry}
\usepackage{babel}
\usepackage{verbatim}
\usepackage{varioref}
\usepackage{amsmath}
\usepackage{amssymb}
\usepackage{graphicx}
\usepackage[unicode=true,pdfusetitle,
 bookmarks=true,bookmarksnumbered=true,bookmarksopen=true,bookmarksopenlevel=1,
 breaklinks=true,pdfborder={0 0 0},backref=false,colorlinks=false]
 {hyperref}

\makeatletter
\newcommand{\code}[1]{\texttt{#1}}

\@ifundefined{date}{}{\date{}}
\makeatother

\begin{document}

\title{\noindent Induction, Coinduction, and Fixed Points:\\
Intuitions and Tutorial{\large{}}\\
\emph{\large{}Using examples from set theory, number theory, and real
analysis}}

\author{Moez A. AbdelGawad\\
\code{moez@cs.rice.edu}}
\maketitle
\begin{abstract}
Recently we presented a concise survey of the formulation of the induction
and coinduction principles, and some concepts related to them, in
five different fields mathematical fields, hence shedding some light
on the precise relation between these fields.

In this article we present few tutorial examples---from set theory,
number theory and real analysis---that illustrate these concepts,
and the intuitions behind them, more concretely.
\end{abstract}

\section{\label{sec:Introduction}Introduction}

Fixed points and concepts related to them, such as induction and coinduction,
have applications in numerous scientific fields. These mathematical
concepts have applications also in social sciences, and even in common
culture, usually under disguise.

These precise formal mathematical concepts are in fact quite common
in everyday life to the extent that in many activities or situations
(or ``systems'') a mapping and a set of its fixed points can be
easily discovered when some little mental effort is exercised. In
particular, in any situation or system where there is some \emph{informal}
notion of repetition or iteration or continual ``come-and-go'' or
``give-and-take'' together with some informal notion of reaching
a stable condition (\emph{e.g.}, equilibrium in chemical reactions
or ``common ground'' between spouses) it is usually the case that
an underlying %
mapping%
, \emph{i.e.}, a mathematical function/map, together with a set of
fixed points of that mapping, can be easily revealed. This %
mapping is sometimes called a `state transition function', and a
fixed point of the mapping is what is usually called a `stable condition'
or a `steady state' (of the ``system'').

These concepts also show up thinly-disguised in visual art and music~\cite{GEB}.
However, applications of these mathematical concepts in scientific
fields, in particular, are plenty and more explicit, including ones
in economics and econometrics~\cite{McLenn2018}, in mathematical
physics~\cite{Penrose2004}, in computer science, and in many other
areas of mathematics itself. Their applications in computer science,
in particular, include ones in programming language semantics---which
we touch upon in this article---as well as in relational database
theory (\emph{e.g.}, recursive or iterated joins) and in concurrency
theory.\footnote{For more details on the use of induction, coinduction, and fixed points
in computer science, and for concrete examples of how they are used,
the reader is invited to check relevant literature on the topic, \emph{e.g.},
\cite{Greiner92,TAPL,Bertot2004,Sangiorgi2012,Kozen2016}.}

Fixed points, induction and coinduction have been formulated and studied
in various subfields of mathematics, usually using different vocabulary
in each field. The interested reader is invited to check our comparative
survey in~\cite{AbdelGawad2018f}. In this article, for pedagogic
purposes, we illustrate the concepts that we presented formally in~\cite{AbdelGawad2018f},
and we attempt to develop intuitions for these concepts using examples
from number theory, set theory and real analysis.

\section{Induction and Coinduction Instances}

\subsection{Induction Instances}

An instance of the set-theoretic induction principle (presented in~$\mathsection$3
of~\cite{AbdelGawad2018f}) is the standard \emph{mathematical induction}
principle. In this well-known instance, induction is setup as follows:
$F$ is the ``successor'' function (of Peano)\footnote{Namely, $F\left(X\right)=\left\{ 0\right\} \cup\left\{ x+1|x\in X\right\} $,
where, \emph{e.g.}, $F\left(\left\{ 0,3,5\right\} \right)=\left\{ 0,1,4,6\right\} $.}, $\mu_{F}$ (the smallest fixed point of the successor function $F$)
is the set of natural numbers $\mathbb{N}$,\footnote{In fact $\mu_{F}$ \emph{defines} the set $\mathbb{N}$ as the set
of all \emph{finite} non-negative whole numbers. Its existence (as
the smallest \emph{infinite} inductive/successor set) is an immediate
consequence of the infinity axiom (and the subset and extensionality
axioms) of set theory (see~\cite[Ch. 4]{Enderton77} and~\cite[$\mathsection$11]{Halmos60}).} and $P$ is any inductive property/set of numbers.
\begin{itemize}
\item An example of an inductive set is the set $P$ of natural numbers
defined as 
\[
P=\{n\in\mathbb{N}|2^{n}>n\}.
\]
(Equivalently, viewing $P$ as a property, its definition expresses
the property that each of its members, say $n$, has its exponential
$2^{n}$ strictly greater than $n$ itself). Set/property $P$ is
an inductive one since $\mathbb{N}\subseteq P$---\emph{i.e.}, $\mu_{F}\subseteq P$---can
be proven \emph{inductively} by proving that $P$ is $F$-closed,
which is sometimes, equivalently, stated as proving that $F$ \emph{preserves
}property $P$ or that property $P$ is \emph{preserved by} $F$.
The $F$-closedness (or `$F$-preservation') of $P$ can be proven
by proving that $F\left(P\right)\subseteq P$ (\emph{i.e.}{\small{},
that} $\forall p\in F\left(P\right),p\in P$). Using the definition
of the successor function $F$, this means proving that $\left(0\in P\right)\wedge\left(\forall x\in\mathbb{N}.x\in P\implies\left(x+1\right)\in P\right)$,
or, in other words, proving that $2^{0}>0$ (\emph{i.e.}, that the
base case, $P\left(0\right)$, holds) and that from $\left(2^{x}>x\right)$
one can conclude that $\left(2^{x+1}>x+1\right)$ (\emph{i.e.},\emph{
}that\emph{ }the inductive case, $P\left(x\right)\implies P\left(x+1\right)$,
holds). This last form is the form of \emph{proof-by-induction} presented
in most standard discrete mathematics textbooks.
\end{itemize}
Another instance of the induction principle is \emph{lexicographic
induction}, defined on lexicographically linearly-ordered (\emph{i.e.},
``dictionary ordered'') pairs of elements~\cite{TAPL,DM}. In~$\mathsection$%
4.1 of~\cite{AbdelGawad2018f} (and~$\mathsection$2.1 of~\cite{AbdelGawad2019b})
we present a type-theoretic formulation of the induction principle
that is closely related to the set-theoretic one. The type-theoretic
formulation is the basis for yet a third instance of the induction
principle---called \emph{structural induction}---that is extensively
used in programming semantics and automated theorem proving (ATP),
including reasoning about and proving properties of (functional) software.

\subsection{Coinduction Instances}

Even though coinduction is the dual of induction, and thus apparently
very similar to it, practical uses of coinduction are relatively obscure
compared to those of induction. For some applications of coinduction
see, \emph{e.g.},~\cite{Brandt1998,Sangiorgi2012,Kozen2016}.

\section{Induction and Coinduction Intuitions}

To develop an intuition for coinduction, let's consider the intuition
behind induction first. While discussing the intuition behind induction
and the goal behind defining sets inductively, Enderton~\cite[p.22]{EndertonLogic72}
states the following (emphasis added)
\begin{quotation}
``We may want to construct a certain subset of a set $U$ by starting
with some initial elements of $U$, and applying certain {[}generating{]}
operations to them \emph{over and over again}. The {[}inductive{]}
set we seek will be the smallest set containing the initial elements
and closed under the operations. Its members will be those elements
of $U$ which can be built up from the initial elements by applying
the operations a \emph{finite} number of times.

{[}...{]}

The idea is that we are given certain bricks to work with {[}\emph{i.e.},
the initial elements{]}, and certain types of mortar {[}\emph{i.e.,
}the generators{]}, and we want {[}the inductive set{]} to contain
just the things we are able to build {[}with a finite amount of bricks
and mortar{]}.''
\end{quotation}
Using this simple intuition we develop a similar intuition for \emph{co}induction.
Like for induction, we want to construct a certain subset of a set
$U$ that includes some initial elements of $U$, and applying certain
generating operations to them over and over again. The \emph{coinductive}
set we seek is now the \emph{largest} set containing the initial elements
and \emph{consistent} under the operations. As such, its members will
be those elements of $U$ which can be built up from the initial elements
by applying the operations a\emph{ finite }or \emph{infinite} number
of times.

The idea, as for induction, is that we are given certain bricks to
work with\emph{ }(\emph{i.e.}, the initial elements) and certain types
of mortar (\emph{i.e.}, the generators), and we want the coinductive
set to contain  \emph{all }things we are able to \emph{fathom} building
(with a finite \emph{or even infinite} amount of bricks and mortar,
but, for consistency, using \emph{nothing else}, \emph{i.e.}, without
using some other kind of building material such as glass or wood).

To illustrate this intuition, a couple of points are worthy of mention:
\begin{itemize}
\item First, while being `\emph{closed} under generating operations' seems
intuitive, being `\emph{consistent} under generating operations'
may initially seem to be unintuitive, and thus be a main hinderance
in having an intuitive understanding of coinduction and coinductive
sets.
\item Second, inductiveness of a set necessitates that elements belonging
to the set result from applying the generating operations (\emph{i.e.},
generators) only a \emph{finite} number of times. This finiteness
condition/restriction is a consequence of induction seeking the \emph{smallest}
subset of $U$ that can be constructed using the generators. As such,
to define a set inductively induction works from below (starts small)
and gets bigger using the generators (which sounds intuitive). Coinduction,
on the other hand, allows, but does not necessitate, applying the
generating operations an infinite number of times. This freedom is
part of coinduction seeking the \emph{largest} subset of $U$ that
can conceivably be constructed using the generators. As such, to define
a set coinductively, coinduction works from above (starts big) and
gets smaller (removing invalid constructions) using the ``generating''
operations---which initially may seem unintuitive (in fact, as explained
precisely below, we will see that the complement of a coinductive
set is a closely-related inductive set, and vice versa%
).
\end{itemize}
The best intuition regarding coinductive sets, however, seems to come
from the fact Kozen and Silva state in~\cite[p.6]{Kozen2016} that
relates coinductive sets to inductive sets and to set complementation
(which, supposedly, are two well-understood concepts):
\begin{quotation}
``It is a well-known fact that a {[}set $X${]} is coinductively
defined as the greatest fixpoint of some monotone operator {[}$F${]}
iff its complement is inductively defined as the least fixpoint of
the dual operator; expressed in the language of $\mu$-calculus, 
\[
\lnot\nu X.F\left(X\right)=\mu X.\lnot F\left(\lnot X\right)
\]
{[}\emph{i.e.}, $\nu X.F\left(X\right)=\lnot\mu X.\lnot F\left(\lnot X\right)${]}.''
\end{quotation}
(Note the \emph{triple} use of complementation. The fact Kozen and
Silva state sounds to be indeed a true and intuitive one. Unfortunately,
however, in~\cite{Kozen2016} Kozen and Silva only stated this fact
in a footnote, merely as a `well-known fact,' without them citing
any references to corroborate the fact).

Trying to understand the intuition behind this well-known fact, which
gives a simple but \emph{indirect} definition of a coinductive set
in terms of an inductive one, gives an intuitive understanding of
coinductive sets; an understanding that depends only on the intuitive
understanding of inductive sets and of (set-theoretic) complementation.

In particular, to gain an intuitive understanding of coinductive sets
using this fact, we focus first on understanding the concept of the
`dual operator' $F^{\delta}$ of a monotone operator $F$ defined
as 
\[
F^{\delta}=\lnot\circ F\circ\lnot
\]
(where $\circ$ is the function composition operator). According to
the definition, the endofunction $F^{\delta}:\mathbb{U}\rightarrow\mathbb{U}$
is an operator on sets (\emph{i.e.}, on subsets of $U$) that, simply,
complements its input set, passes that complement to generator $F$,
then returns the complement of what $F$ returns as its own result
set.\\
Given that $F$ is assumed monotonic, $F^{\delta}$ is easily proved
to be monotonic too. That is, we have
\[
\forall X,Y\subseteq U.X\subseteq Y\implies F^{\delta}\left(X\right)\subseteq F^{\delta}\left(Y\right),
\]
because (by the contravariance of complementation\footnote{Also called \emph{antimonotonicity} or \emph{antitonicity} of $\lnot$,
\emph{i.e.}, $\forall A,B\subseteq U.A\subseteq B\implies\lnot B\subseteq\lnot A$.}) we have $X\subseteq Y\implies\lnot Y\subseteq\lnot X$ which (by
the monotonicity of $F$) implies that $F\left(\lnot Y\right)\subseteq F\left(\lnot X\right)$,
which implies (again, by complementation contravariance) that $F^{\delta}\left(X\right)=\lnot F\left(\lnot X\right)\subseteq\lnot F\left(\lnot Y\right)=F^{\delta}\left(Y\right)$
as required. As such, like $F$, $F^{\delta}$ is also a sets-generator.\\
Now, as a sets generator $F^{\delta}:\mathbb{U}\rightarrow\mathbb{U}$
has a least fixed point, called $\mu_{F^{\delta}}$, which is an \emph{inductively}
defined set. The complement of set $\mu_{F^{\delta}}$ (\emph{i.e.},
the set $\lnot\mu_{F^{\delta}}=U-\mu_{F^{\delta}}$) is the \emph{coinductively}
defined set we seek, \emph{i.e.}, is the the \emph{greatest} fixpoint
of $F$ (\emph{i.e.}, $\nu_{F}$). As such, eliding the composition
operator (as is customary), we have
\begin{equation}
\nu_{F}=\lnot\mu_{\lnot F\lnot}\label{eq:coind-ind-neg}
\end{equation}
which, we believe, expresses precisely and concisely the intuition
behind coinduction and coinductive sets.

\subsubsection{Notes}
\begin{itemize}
\item Noting that complementing a finite subset of an infinite set $U$
produces a \emph{cofinite} set (`the complement of a finite set'),
we can intuitively see that when starting with a finite set (the initial
``building bricks,'' using Enderton's terminology) then applying
complementation \emph{thrice} (as is done in the definition of $\nu_{F}$)
defines some special sort of an infinite set (particularly, some special
sort of a cofinite set). This special sort of infinite sets constitutes
coinductive sets.
\item For example, given the Peano generator $F\left(X\right):\wp\left(\mathbb{N}\right)\rightarrow\wp\left(\mathbb{N}\right)=\left\{ 0\right\} \cup\left\{ x+1|x\in X\right\} $
that we discussed earlier, we have 
\[
F^{\delta}\left(X\right)=\mathbb{N}-\left(\left\{ 0\right\} \cup\left\{ x+1|x\in\left(\mathbb{N}-X\right)\right\} \right).
\]
As such, if $\mathbb{N}^{+}$ is the set of positive integers, $\mathbb{N}_{e}$
is the set of even natural numbers, and $\mathbb{N}_{o}$ is the set
of odd natural numbers, then (as readers are invited to ascertain
for themselves) we have 
\[
F^{\delta}\left(\phi\right)=\phi,\,F^{\delta}\left(\mathbb{N}\right)=\left\{ 1,2,3,...\right\} =\mathbb{N}^{+},\,F^{\delta}\left(\{0,1,2\}\right)=\left\{ 1,2,3\right\} ,\,F^{\delta}\left(\left\{ 1,2\right\} \right)=\left\{ 2,3\right\} ,
\]
\[
F^{\delta}\left(\mathbb{N}_{o}\right)=\mathbb{N}_{e}-\left\{ 0\right\} ,\,F^{\delta}\left(\mathbb{N}_{e}\right)=\mathbb{N}_{o},\textrm{ and }F^{\delta}\left(\mathbb{N}^{+}\right)=\mbox{\ensuremath{\mathbb{N}} -}\left\{ 0,1\right\} =\mathbb{N}^{++},
\]
which intuitively justifies stating that $F^{\delta}\left(X\right)=\left\{ x+1|x\in X\right\} $.\footnote{Note that for all $X$, $F^{\delta}\left(X\right)$ never contains
0 (\emph{i.e.}, the initial ``bricks''), and also that $F^{\delta}\left(\mathbb{N}\right)=\mathbb{N}^{+}$
and $F^{\delta}\left(\mathbb{N}^{+}\right)=\mathbb{N}^{++}$ are \emph{proper}
subsets of $\mathbb{N}$. Nevertheless, $F^{\delta}$ is monotonic---which
initially appears to defy intuition, as is usual when reasoning about
\emph{infinite} sets.}\\
As is obvious from $F^{\delta}\left(\phi\right)$, the set $\phi$
is the least fixed point of $F^{\delta}$, and thus $\mu_{F^{\delta}}=\phi$.
Accordingly, we have 
\[
\nu_{F}=\mathbb{N}-\mu_{F^{\delta}}=\mathbb{N}-\phi=\mathbb{N}.
\]
This means that, for this particular $F$ (with the chosen domain
$\wp\left(\mathbb{N}\right)$), we have $\mu_{F}=\nu_{F}$ (\emph{i.e.},
the least fixed point and the greatest fixed point of $F$ agree,
and $F$ has only one fixed point). %

\end{itemize}
\medskip{}

To further strengthen our intuitions, let's consider yet a third
intuition about coinductive sets---the one offered in~\cite[p.61]{Forster2003},
which states that
\begin{quote}
``{[}An element{]} will belong to a coinductive {[}set{]} as long
as there is \emph{not} a good finite reason for it \emph{not} to.''\end{quote}
\begin{quotation}
which is also expressed, a little less-precisely, in~\cite[p.2]{Kozen2016}
as\end{quotation}
\begin{quote}
``A property {[}\emph{e.g.}, an element belonging to a set{]} holds
by induction if there is good reason {[}\emph{e.g.}, a finite construction
of the element{]} for it to hold; whereas a property holds by coinduction
if there is \emph{no} good reason {[}\emph{e.g.}, a finite construction{]}
for it \emph{not} to hold.''
\end{quote}
Connecting these two similar intuitions with the prior intuition (\emph{i.e.},
the one based on Kozen and Silva's well-known fact), we can see that
the role of the dual operator $F^{\delta}$ (\emph{i.e.}, in defining
$\nu_{F}$) is to define (\emph{inductively}) those elements for which
there is ``a good finite reason'' to \emph{not} belong to $\nu_{F}$
(we call these the \emph{rejected} elements). $F^{\delta}$ defines
these rejected elements by precomposing $F$ with negation (set complementation)
then postcomposing the resulting operator with negation again. The
role of $\mu_{F^{\delta}}$, the least fixed set of $F^{\delta}$,
is then to collect \emph{all} these rejected elements (all of which
have a finite ``bad'' reason---\emph{i.e.}, inconsistency with $F$---to
belong, or, equivalently, a finite good reason \emph{not} to belong),
and \emph{nothing but} \emph{them} (since $\mu_{F^{\delta}}$ is a
minimal inductive set), in one set. Then, finally, the role of the
third (\emph{i.e.}, the outermost) negation in the definition of $\nu_{F}$
is to exclude those rejected elements from belonging to $\nu_{F}$
(the greatest fixed set of $F$).

\subsubsection{Notes}
\begin{itemize}
\item Combining the intuitive understanding of inductive and coinductive
sets, it is now easy to observe that a sets-generator $F$ defined
over (subsets of) an infinite universe $U$ defines three\emph{ }disjoint
(and possibly empty) subsets of $U$:

\begin{enumerate}
\item A set $\mu_{F}$ containing the ``good and finite'' elements of
$U$ (relative to $F$) that ``break no rules'' (\emph{i.e.}, that
can be constructed by $F$, and are thus consistent with it), and
that can be constructed finitely (in a finite number of steps/applications
of $F$),
\item A set $\nu_{F}-\mu_{F}$ (note that $\mu_{F}\subseteq\nu_{F}$ always
holds) containing the ``in-between, good but infinite'' elements
of $U$ (relative to $F$) that also break no rules, (\emph{i.e.},
can be constructed by $F$ and are thus consistent with it), but that
cannot be finitely constructed by it (and thus do not belong to $\mu_{F}$
but only to $\nu_{F}$), and
\item A set $U-\nu_{F}$ containing the ``bad'' elements of $U$ (relative
to $F$) that ``break the rules'' (of $F$) and thus cannot be finitely
or even infinitely constructed by $F$\emph{ }(\emph{i.e.}, are inconsistent
with $F$) and thus are not included even in $\nu_{F}$ (these are
elements of $U$ that, whether finite or infinite elements, cannot
be constructed by $F$, whether $F$ is applied a finite or infinite
number of times, and are thus ``totally inconsistent'' with $F$).
\end{enumerate}

This observation can be summarized as saying that relative to a generator
$F$ elements of set $U$ are either: \emph{finitely-consistent} with
$F$, \emph{infinitely-consistent} with $F$, or (\emph{finitely or
infinitely})\emph{ inconsistent} with $F$.

\item An example of dual operators are the logical operators $\forall$
(forall) and $\exists$ (there exists) in first-order logic. Note
that $\forall\equiv\lnot\exists\lnot$ and $\exists\equiv\lnot\forall\lnot$.
Same applies for $\cap$ and $\cup$ (in set theory), and for $\wedge$
and $\vee$ (in propositional logic). The statements that these pairs
of operators are dual operators are usually called `De Morgan's
Laws.'%

\item Note that `complementation as negation' makes sense in set theory
(and, accordingly, also in structural type theory and in first-order
logic. See~$\mathsection\mathsection$%
4.1 and~%
5 of~\cite{AbdelGawad2018f}), thus making coinductive sets (and
coinductive structural data types and coinductive first-order logical
statements) relatively simple to understand, \emph{i.e.}, are understandable
based on the intuitive understanding of inductive sets. A counterpart
of set-theoretic complementation may not, however, be easily definable
as the meaning of negation ($\lnot$) in other mathematical fields~\cite{Horn2001}.
This makes coinductive objects in these fields a bit harder to reason
about. (We intend to deliberate on this point more in future versions
of this article.%
{} But see also~$\mathsection$%
3.2 of~\cite{AbdelGawad2019b} for a discussion of type negation
in the context of OO/nominal type theory).
\item Another less-obvious instance of a coinductive set is the standard
subtyping relation in (nominally-typed) object-oriented programming
languages. See the notes of~$\mathsection$2.2 of~\cite{AbdelGawad2019b}
for more details.
\end{itemize}

\section{Illustrating Induction and Coinduction}

In this section we strengthen our intuitive understanding of pre-fixed
points, post-fixed points, and fixed points by illustrating these
concepts in set theory, and also by presenting a pair of examples
(and exercises) from analysis and number theory---examples that, as
such, are most likely familiar to many readers---that exemplify these
concepts.

\subsection{In Set Theory}

\begin{figure*}
\noindent \begin{centering}
\includegraphics[scale=0.8]{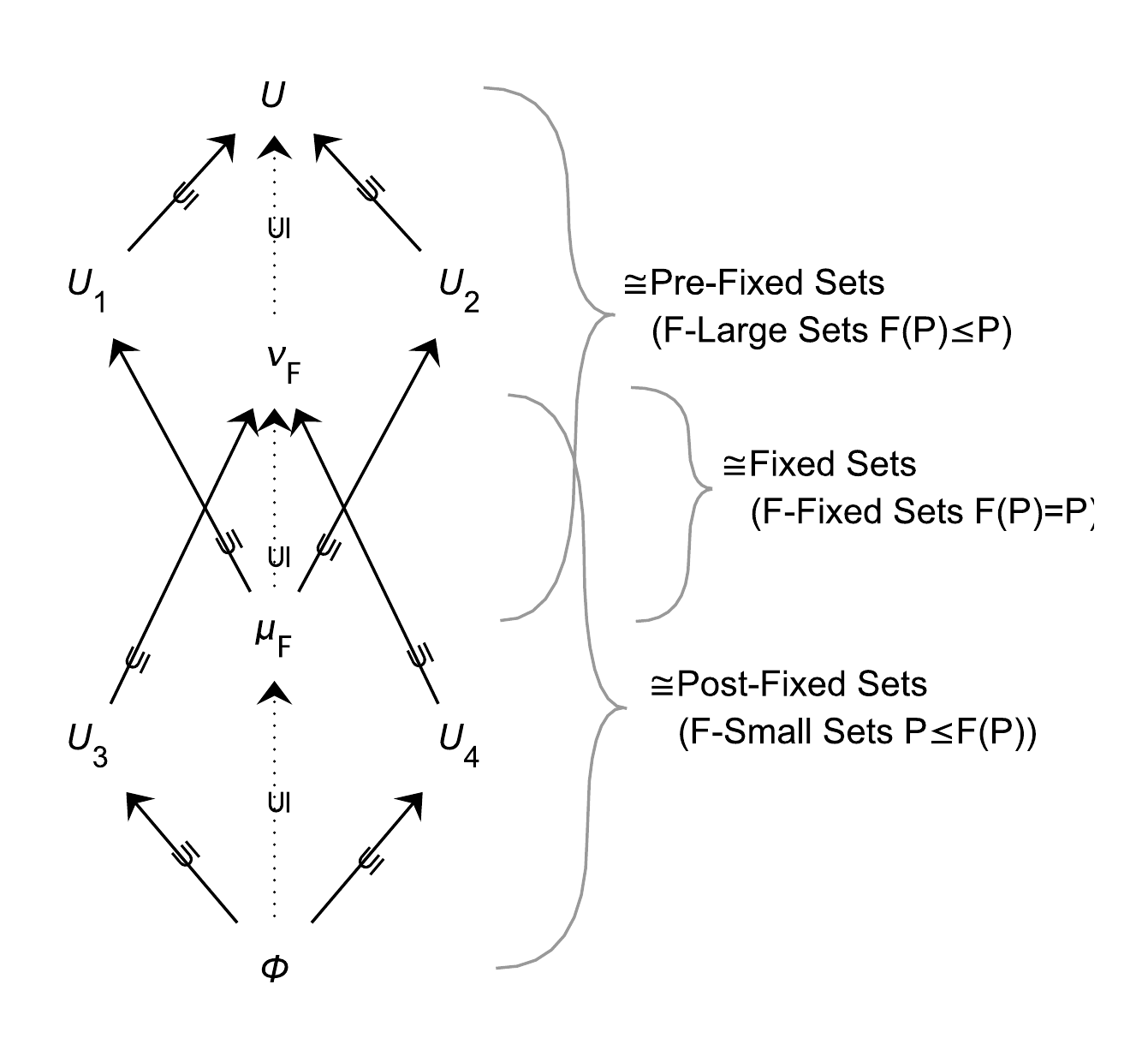}
\par\end{centering}

\protect\caption{\label{fig:Illus}Illustrating the pre-fixed points, post-fixedpoints,
and fixed points of a generator (\emph{i.e.}, monotonic function)
$F:\mathbb{U}\rightarrow\mathbb{U}$ where $\mathbb{U}=\wp(U)$.}
\end{figure*}

Figure~\vref{fig:Illus} visually illustrates the main concepts we
discuss in this article in the context of set theory, by particularly
presenting the pre-/post-/fixed points of some abstract generator
$F$ over the powerset (a complete lattice) $\mathbb{U=}\wp\left(U\right)$
for some abstract set $U$.

Figure~\ref{fig:Illus} illustrates that subsets $U_{1}$, $U_{2}$
and $\mu_{F}$ of $U$ are among the $F$-closed/$F$-large/$F$-pre-fixed
subsets of $U$ (\emph{i.e.}, the upper diamond, approximately), while
subsets $U_{3}$, $U_{4}$ and $\nu_{F}$ are among its $F$-consistent/$F$-small/$F$-post-fixed
subsets (\emph{i.e.}, the lower diamond, approximately). Of the sets
illustrated in the diagram, only sets $\mu_{F}$ and $\nu_{F}$ are
$F$-fixed subsets of $U$, \emph{i.e.}, belong to the intersection
of $F$-pre-fixed and $F$-post-fixed subsets (\emph{i.e.}, the inner
diamond, exactly). It should be noted that some other subsets of $U$
(\emph{i.e.}, points/elements of $\mathbb{U}$) may neither be among
the pre-fixed subsets of $F$ nor be among the post-fixed subsets
of $F$ (and thus are not among the fixed subsets of $F$ too). Such
subsets are not illustrated in Figure~\ref{fig:Illus}. (If drawn,
such subsets would lie, mostly, outside all three diamonds illustrated
in Figure~\ref{fig:Illus}.)

Figure~\ref{fig:Illus} also illustrates that $\mu_{F}\leq\nu_{F}$
is true for any set $U$ and \emph{any} generator $F$. However, depending
on the particular $F$, it may (or may not) be the case that $\mu_{F}=\nu_{F}$
(\emph{e.g.}, if $F$ is a constant function or if $F$ happens to
have only one fixed point).\footnote{In the context of category theory, the symbol $=$ is usually interpreted
as denoting an \emph{equivalence} or \emph{isomorphism} relation between
objects, rather than denoting the equality relation. See $\mathsection$6
of~\cite{AbdelGawad2018f}.}

With little alterations (such as changing the labels of its objects,
the direction of its arrows, and the symbols on its arrows) the diagram
in Figure~\ref{fig:Illus} can be used to illustrate these same concepts
in the context of order theory, type theory, first-order logic or
category theory. (Exercise: Do that.)

\subsection{In Analysis}

Most readers will definitely have met (and have been intrigued by?)
fixed points in their high-school days, particularly during their
study of analysis (the subfield of mathematics concerned with the
study of real numbers $\mathbb{R}$%
{} and with mathematical concepts related to them). To further motivate
the imagination and intuitions of readers, we invite them to consider
the function $f:\mathbb{R\rightarrow\mathbb{R}}$ defined as 
\[
f\left(x\right)=\left(x-5\right)^{3}-5x+29
\]
and graphed in%
\begin{figure*}
\noindent \begin{centering}
\includegraphics[scale=0.4]{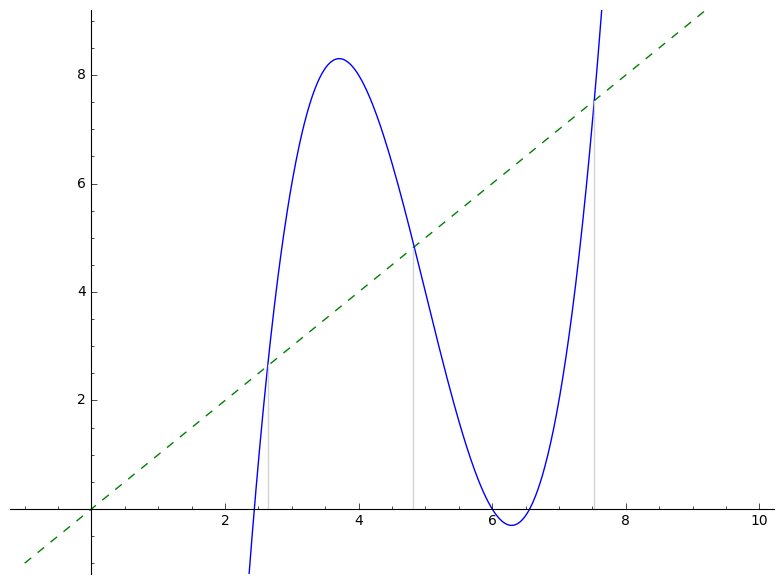}
\par\end{centering}

\protect\caption{\label{fig:fixed-zeros}Graph of function $f\left(x\right)$.}
\end{figure*}
 Figure~\vref{fig:fixed-zeros}. In particular, we invite the reader
to decide: (1) which points of the \emph{totally-ordered }real line
(the \emph{x}-axis) are pre-fixed points of $f$, (2) which points/real
numbers are post-fixed points of $f$ and, (3) most easily, which
points are fixed points of $f$.\footnote{To decide these sets of points using Figure~\ref{fig:fixed-zeros},
readers should particularly take care: (1) not to confuse the fixed
points of $f$ (\emph{i.e., }crossings of the graph of $f$ with the
graph of the identity function, which are solutions of the equation
$f\left(x\right)-x=0$) with the \emph{zeroes} of $F$ (\emph{i.e.},
crossings with the $x$-axis, which are solutions of the equation
$f\left(x\right)=0$), and (2) not to confuse pre-fixed points of
$f$ for its post-fixed points and vice versa.\\
Since $F$ is not a monotonic function\emph{ }(\emph{i.e.}, is not
a generator) we compensate by giving some visual hints in Figure~\ref{fig:fixed-zeros}
so as to make the job of readers easier. The observant reader will
recall that diagrams such as that in Figure~\ref{fig:fixed-zeros}
are often used (\emph{e.g.}, in high-school/college math textbooks),
with ``spirals'' that are zeroing in on fixed points overlaid on
the diagrams, to explain iterative methods---such as the renowned
Newton-Raphson method---that are used in numerical analysis to compute
numerical derivatives of functions in analysis. These methods can
be easily seen to be seeking to compute fixed points of some related
functions. As such, these numerical analysis methods for computing
numerical derivatives can also be explained in terms of pre-fixed
points and post-fixed points.}\\
(Exercise: Do the same for the \emph{monotonic} function $g:\mathbb{R}\rightarrow\mathbb{R}$
defined as
\[
g\left(x\right)=\begin{cases}
\dfrac{x^{2}}{4} & 0\leq x<4\\
x & 4\leq x\leq6\\
\dfrac{(x-8)^{3}}{4}+8 & 6<x\leq10
\end{cases}
\]
and depicted in%
\begin{figure*}
\noindent \begin{centering}
\includegraphics[scale=0.4]{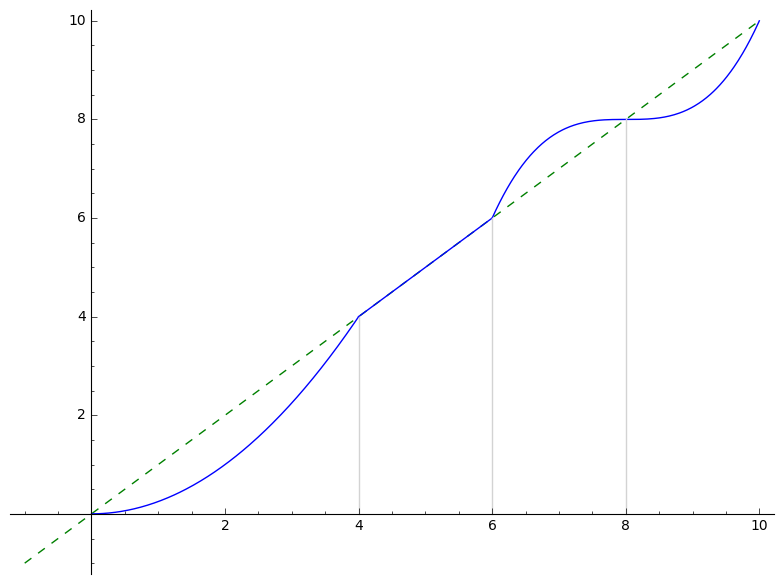}
\par\end{centering}

\protect\caption{\label{fig:fixed-set}Graph of monotonic function $g\left(x\right)$.}
\end{figure*}
 Figure~\vref{fig:fixed-set}, and then relate your findings to Figure~\ref{fig:Illus}%
. Now also relate your findings regarding $f\left(x\right)$ in Figure~\vref{fig:fixed-zeros}---or
regarding mono\-tonic/mono\-tonically-increasing and anti\-monotic/mono\-tonically-decreasing
sections of $f\left(x\right)$---to Figure~\ref{fig:Illus}. Based
on your relating of Figure~\ref{fig:fixed-zeros} and Figure~\ref{fig:fixed-set}
to Figure~\ref{fig:Illus}, what do you conclude?).

The main goal of presenting these two examples from analysis is to
let the readers strengthen\emph{ }their understanding of pre-/post-/fixed
points and dispel any misunderstandings they may have about them,
by allowing them to connect examples of fixed points that they are
likely familiar with (\emph{i.e.}, in analysis) to the more general
but probably less familiar concepts of fixed points found in order
theory, set theory, and other branches of mathematics. Comparing the
illustration of fixed points and related concepts in set theory (\emph{e.g.},
as in Figure~\ref{fig:Illus}) with illustrations of them using examples
from analysis (\emph{e.g.}, as in Figure~\ref{fig:fixed-zeros} and
Figure~\ref{fig:fixed-set}) makes evident the fact that a function
over a partial order (such as $\mathbb{U}$) is not as simple to imagine
or draw as doing so for a function over a total order (such as $\mathbb{R}$)
is. However, illustrations and examples using functions over total
orders can be too specific and thus misleading, since they may hide
the more generality, the wider applicability, and the greater mathematical
beauty of the concepts depicted in the illustrations.

\subsection{In Number Theory}

For a non-visual example, consider perfect numbers in arithmetic.\footnote{This example is taken from~\cite[p.5]{Enderton77}. The example,
being tangential and in a different context than ours, is mentioned
\emph{without} reference to pre-/post-/fixed points. More on perfect
numbers can be found in other standard texts on arithmetic and number
theory.} A positive integer is \emph{perfect} if it equals the sum of its
smaller divisors, \emph{e.g.}, 6=1+2+3. It is \emph{deficient} (or
\emph{abundant}) if the sum of its smaller divisors is less then (or
greater than, respectively) the number itself. The first four perfect
numbers are 6, 28, 496, and 8128.

A watchful reader will immediately note that deficient numbers are
the proper pre-fixed points (ones that are not fixed points) of a
function from numbers to numbers, abundant numbers are the proper
post-fixed points of the function, and perfect numbers are its fixed
points. That function, namely, is the endofunction that sums the smaller
divisors of its input number.

Lets call this function $sd$ (for `sum of divisors'). If $\mathbb{P}$
is the (totally-ordered) set of positive whole numbers, then deficient
numbers are exactly the members of the set $dn=\left\{ p\in\mathbb{P}|sd\left(p\right)<p\right\} $
of proper pre-fixed points of $sd$, abundant numbers are exactly
members of the set $an=\left\{ p\in\mathbb{P}|p<sd\left(p\right)\right\} $
of proper post-fixed points of $sd$, and perfect numbers are exactly
members of the set $pn=\left\{ p\in\mathbb{P}|p=sd\left(p\right)\right\} $
of fixed points of $sd$. (Note that $sd$ is \emph{not} a mono\-tonic/co\-var\-iant/in\-creas\-ing/mono\-tonically-in\-creasing
function, and that $\mathbb{P}$ is \emph{not} a complete lattice.
However, the notions pre-fixed points, post-fixed points, and fixed
points still have relevance in such a context. See~$\mathsection$%
2.2 of~\cite{AbdelGawad2019b} for a similar situation.)

\bibliographystyle{plain}

\end{document}